\documentclass[12pt,dvips]{article}

\usepackage{amsmath,amssymb,exscale}
\usepackage{array,multicol}
\usepackage{afterpage,float,flafter}
\usepackage{epsfig,rotating,pifont}
\usepackage{cite}
\makeatletter
\@addtoreset{equation}{section}
\makeatother

\title{
\vspace*{-2.5cm}
\begin{flushright}
\end{flushright}
\vspace{1cm} 
\Large\textbf{Fat Euclidean Gravity \\
with \\
 Small Cosmological Constant}
\vspace*{.5cm}
\author{\large \textbf{
Raman Sundrum\footnote{email: ~ sundrum@pha.jhu.edu} 
}\\
\\
\emph{
Department of Physics and Astronomy} \\ 
\emph{Johns Hopkins University} \\ 
\emph{3400 North Charles St}. \\ 
\emph{Baltimore, MD 21218-2686}}}

\date{}
\begin{document}
\maketitle
\thispagestyle{empty}
\vspace*{-0cm}
  
\begin{abstract} 
The cosmological constant problem is usually considered an inevitable 
feature of any effective theory capturing well-tested gravitational and 
matter physics, without regard to the details of short-distance 
gravitational couplings. In this paper, 
a subtle effective description avoiding the problem 
is presented in a first quantized language, 
consistent with experiments and the Equivalence Principle. First quantization 
allows a minimal domain of validity to be carved out by 
cutting on the proper length of particle worldlines. This is facilitated 
by working in (locally) Euclidean spacetime, although considerations of 
unitarity are still addressed by analytic continuation from Lorentzian 
spacetime. The new effective description demonstrates that the 
cosmological constant problem {\it is} sensitive to short-distance details of 
gravity, which can be probed experimentally. 
``Fat Gravity'' toy models are presented, 
illustrating how gravity might shut off at short but testable distances, 
in a generally covariant manner that 
 suppresses the cosmological constant.  This paper improves 
on previous work by allowing generalizations to massless matter, non-trivial 
spins, non-perturbative phenomena, and multiple (metastable) vacua.
\end{abstract} 
  
\newpage 
\renewcommand{\thepage}{\arabic{page}} 
\setcounter{page}{1}


\section{Introduction}

Imagine integrating out all physics above a TeV to arrive at an effective 
field theory of  gravity and Standard Model 
matter, 
\begin{equation}
\label{teveft}
{\cal Z} = \int [{\cal D} g_{\mu \nu}]_{_{p_{\mu} < {\rm TeV}}} 
[{\cal D} \psi_{SM}]_{_{p_{\mu} < {\rm TeV}}}~  e^{i\int d^4x ~{\cal L}_{eff}}.
\end{equation}
This effective theory beautifully accounts for everything we have tested 
experimentally in fundamental physics, while obeying the sacred 
principles of Equivalence, relativity, unitarity, 
and locality (down to $1/$TeV). 
However, to match the real world 
the theory requires extreme fine-tuning in the 
cosmological constant \cite{weinberg}.  
So it is probably wrong. 

In looking for what to change, we see that there is a vast tract of 
completely unexplored gravitational physics between $10^{-3}$ eV 
$(\sim (100$ microns$)^{-1})$ \cite{gravexpt1} and 
a TeV which has, by default in Eq. (\ref{teveft}),
 been assumed to be a desert. Could new gravitational physics in this 
regime eliminate cosmological constant fine-tuning? A reasonably 
model-independent way of approaching this is to imagine integrating out 
just the gravitational physics in this regime, yielding
\begin{equation}
\label{2cutoff}
{\cal Z} = \int [{\cal D} g_{\mu \nu}]_{_{10^{-3} {\rm eV}}} 
[{\cal D} \psi_{SM}]_{_{\rm TeV}}~  e^{i\int d^4x ~{\cal L}_{eff}}.
\end{equation}
This effective description has the following properties: 

i) It economically accounts for everything we have 
experimentally tested.

ii) It is fully relativistic. 

iii) It obeys the Equivalence Principle.

iv) The purely matter couplings are manifestly local down to 
distances of order $1/$TeV. 

v) The couplings of gravity to itself and to matter are manifestly 
local down to distances of  

~ $1/10^{-3}$eV 
$\sim 100$ microns.

vi)  It is unitary in the regime of small momentum 
transfers $< 10^{-3}$eV. 

vii) It is unitary up to a TeV in the pure matter sector, when gravitational
interactions are neglected.

However Eq. (\ref{2cutoff}) does have its price. 
When matter coupled to gravity is considered 
 above $10^{-3}$ eV, unitarity is violated. 
To see this, recall that we are positing new gravitational physics far 
below a 
TeV. There is plenty of phase space 
for a process such as
\begin{equation}
e^+ ~ e^- \rightarrow_{\sqrt{s} = 10 {\rm GeV}} ~ e^+ ~ e^- + {\rm 
new ~gravitational ~stuff},
\end{equation}
where states within the domain of  Eq. (\ref{2cutoff}) 
are taken to states outside the domain. 
In the fundamental theory there should be some probability for this to 
happen, and therefore there must be a violation of unitarity within 
Eq. (\ref{2cutoff}). On this basis of a lack of what we might call 
``kinematic closure'', a purist would disqualify Eq.(\ref{2cutoff})
as an effective field theory. 

 Of course the loss of probability in Eq.(\ref{2cutoff}) need only be of 
gravitational strength, $G_{N}.$TeV$^2$, which is far below our 
experimental precision. 
We only see gravitational effects 
experimentally because
the smallness of $G_{N}$ is compensated by energies very
 much larger than 
a TeV, arranged by having 
sources containing enormous numbers of fundamental particles. Such sources 
are necessarily large and slowly varying
 and therefore the relevant gravitational 
degrees of freedom have predominantly large wavelengths, captured by 
Eq. (\ref{2cutoff}). Thus there is a domain of validity for 
Eq. (\ref{2cutoff}), encompassing all experiments and observations 
performed so far  and
 with all the advantages  of effective 
field theory, as long as we are willing to neglect $G_{N}.$TeV$^2$
when uncompensated by macroscopic factors. 

This ``almost effective field theory'' is a good starting 
point for studying the cosmological constant problem. It enjoys the 
properties of effective field theory to within experimental precision. 
Furthermore, after decades of looking maybe there is simply no full effective 
field theory in which the cosmological constant can be understood
because the resolution lies outside of field theory, even at low energy. 
 Of course, the highest standard for any resolution is embedding within some 
fundamental theory of gravity and matter, but perhaps this need not be the 
first step.

Alas, the rather minimal description of the world given by Eq. (\ref{2cutoff}) 
still does not evade cosmological constant fine-tuning. 
In Feynman diagram language the
dominant
diagrams contributing to the cosmological constant are those with 
very soft  gravitational fields on external lines and hard 
matter fields and  couplings in the interior. All of these ingredients are 
still present in Eq. (\ref{2cutoff}). Furthermore, it seems that there is now
little room left for new physics to be hiding that might solve the problem. 
We apparently have 
already pared down the effective description to just encompass the 
experimentally tested regimes.  It is this argument, 
in some form or another,  
that convinces many 
theorists that there can be no new local physics that resolves the 
cosmological constant problem. 

Nevertheless,
in earlier papers \cite{97} \cite{old} I have argued that  
Eq. (\ref{2cutoff}) is not minimal enough, that it implicitly makes an 
untested assumption about the short distance coupling of 
gravity to matter, and that there is a different 
possibility, ``fat gravity'',
which eliminates the cosmological constant problem. 
Fat gravity has a distinct and exciting prediction for short-distance
 tests of gravity \cite{gravexpt1} \cite{gravexpt2} \cite{ann}, 
namely a suppression of the gravitational force. 

Some related ideas and issues in the literature are as follows.
The general possibility of a 
connection between the cosmological constant problem and sub-millimeter 
gravity was pointed out many years ago in Ref. \cite{banks}. 
Failed attempts to explicitly realize cosmological constant relaxation 
mechanisms based on light scalars are reviewed in Ref. \cite{weinberg}.
Ref. \cite{beane} 
argued that {\it any} resolution of the 
cosmological constant problem within local effective field theory would 
yield light scalars with sub-millimeter range, which  should 
be sought experimentally. Ref. \cite{moffat} described an approach to the 
cosmological constant problem with sub-millimeter non-locality in the 
gravitational couplings. While this shares some similarities with fat 
gravity, there are also important qualitative differences which make 
satisfying the Equivalence Principle problematic. 
Ref. \cite{gia} described an extra-dimensional 
approach to the cosmological constant problem which invoked fat gravity as a 
natural corollory of a sub-millimeter higher-dimensional Planck length.

 In this paper I want to rederive the fat-gravity 
loop-hole in two distinct steps:

I) Replace  Eq. (\ref{2cutoff}) with an effective description with  the 
same virtues (i -- vii), but with naturally small cosmological constant.
This step is obviously powerful, but nevertheless conservative. I do not 
 posit what the new gravitational physics above $10^{-3}$ eV is, I just make 
manifest the subtle loop-hole in the cosmological constant problem 
that is already present subject to only (i -- vii). 

II) Present a toy model of fat gravity, that is actually commit to what 
the gravitational physics  above $10^{-3}$ eV is like, 
such that the cosmological 
constant is indeed small. In fact this step requires only a reinterpretation 
of the results of step I, the mathematics is the same. The fat 
gravity model is useful in 
that it is simple and you can play with it, push it around and ask some
tough diagnostic questions. You can also see qualitatively what happens in 
short distance tests of gravity. But it is a toy because although the 
graviton is an extended object, with  a size $\sim 100$ microns, 
the relativistic corollory, graviton excitations at $10^{-3}$ eV,
have not been included. Associated with their absence
there are Planck-suppressed 
violations of unitarity, for gravitational momentum transfers 
$\sim 10^{-3}$ eV. Well above this scale unitarity is trivially satisfied
simply because fat gravity does not contain large gravitational momentum 
transfers, and well below this scale unitarity is non-trivially satisfied.
It is a rather satisfying part of the model that it gets this 
low-energy unitarity right.

Why is a 
 fully-fledged fat gravity model with graviton excitations so hard to build?
Because it requires a relativistic theory of an extended 
object, one of whose modes is the massless graviton. Therefore it is at 
least as tough a venture as string theory.
 But  a fat gravity phase of string theory (with 
relatively point-like matter) is still unknown. (Neither is there known 
to be a general No-Go theorem.) 
Nevertheless I presume that if a fully unitary and relativistic fat gravity 
exists it is probably  within some as yet unknown phase of string theory.

There are two central reasons why the loop-hole I want to point out is easily 
missed. The first is that we are used to thinking about the coarse-graining
procedure  
(in order to get effective field theories from more fundamental ones) 
in terms of momentum (or other 
de-localized) modes. The subtlety in the present problem is simple to see in 
position space but appears like a conspiracy in momentum space. Therefore 
I will do the  coarse-graining in position space. This is most easily done in 
Euclideanized field theory, so I will work in locally Euclidean 
spacetime. Nevertheless, the statements about unitarity, (vi) and (vii), 
 will be recovered by 
analytic continuation from Lorentzian spacetime. 

A second reason is that 
the usual formalism of second quantization, and associated Feynman diagrams 
of Eq. (\ref{2cutoff}),
 autmatically correlate certain effects in a way which 
would be justified if there were a single cutoff, as in Eq. (\ref{teveft}), 
but not if there are unknown new degrees of freedom above the lower 
cutoff, as in Eq. (\ref{2cutoff}). 
 For this reason we 
will begin our discussion by treating matter in first quantized effective 
theory. Once we have learned the basic lesson,  we will generalize back to 
the more familiar second quantized formalism.  
Strictly speaking, steps I and II of this paper are carried out subject to 
all conditions (i -- vii) for perturbative matter comprised only of scalar 
fields with general couplings and masses, coupled to gravity. 
However, a more powerful ``block-spin'' \cite{block-spin} 
approach to coarse-graining
is also presented which suggests a clear generalization to non-perturbative 
and general matter. The price for the latter approach
 is that the cutoff on gravitational 
physics, $10^{-3}$ eV, is not manifestly Lorentz invariant (and hence 
not manifestly generally covariant), but yields 
Lorentz invariance order by order in a  soft-graviton expansion. 
I believe that this 
feature is a purely technical rather than a conceptual obstacle, because 
there is no such issue for the restricted case of 
perturbative scalar matter. Purely matter 
interactions are manifestly 
Lorentz invariant. 

Finally, after going through the 
detailed derivations and understanding their point, 
there is a  simple and manifestly
generally covariant (and Lorentz invariant) prescription
 enforcing the results 
of the derivations in the fully realistic setting which avoids the 
cosmological constant fine-tuning. Without the derivations of this 
paper the presciption might appear ad hoc. This
 prescription was anticipated in 
Ref. \cite{old}.

The paper is organized as follows. In Section 2, the technical problem of 
the unboundedness of the gravitational action in Euclidean space 
\cite{hawking} is discussed,
including how the problem is avoided in perturbative gravity, which is 
all we consider here. We also discuss  UV regulators 
to cut off gravity at $\sim 10^{-3}$ eV. In Section 3, 
we introduce scalar matter in first quantized formalism. In Section 4, 
we show that for scalar matter there is an even more minimal effective 
description than Eq. (\ref{2cutoff}) which continues to satisfy (i -- vii) 
but has naturally small cosmological constant. 
 The same mathematics is then turned into a  
toy fat gravity model. In Section 5, we discuss how 
the short distance static gravitational force is necessarily suppressed 
relative to Newton's Law in the fat gravity model and how the transition 
distance scale is narrowly constrained by present experiments and 
cosmological constant naturalness.
 In Section 6, we discuss a different 
cutoff procedure that gives up manifest Lorentz invariance 
but makes the generalization to general 
non-perturbative second quantized matter much more obvious, and retains the 
small cosmological constant. We first present this in a simpler system where 
not just matter, but even gravity is replaced by a weakly coupled 
scalar field. This step is not strictly necessary but makes for easier 
reading.  
In Section 7, we repeat Section 6 but now for real gravity, and deal 
with the fact that the cutoff procedure does not manifestly preserve 
general coordinate invariance. We are careful to distinguish the steps 
needed to recover this symmetry in the infrared from the cosmological 
constant problem. 
In Section 8, we generalize our results for real gravity 
coupled to second quantized scalar matter to the case of general second 
quantized matter such as the Standard Model. Thus issues such as the 
non-perturbative QCD effects on the cosmological constant can be 
easily dealt with.
In Section 9, 
we study the case when there are multiple vacua in the matter sector and 
show that within our fat gravity model it is the true vacuum which has 
small cosmological constant while false vacua have large positive 
cosmological constants. In Section 10 we give a simple prescription which is 
manifestly generally coordinate invariant for practical
computations in the Standard Model
coupled to soft gravity, and which encapsulates the lessons of this paper and 
avoids cosmological constant fine-tuning. The prescription depends on 
the preceding sections for its justification. 
Section 11 provides conclusions. A technical 
proof appears in the appendix.

I imagine a reader, jaded by having witnessed many unsatisfactory attempts on 
the cosmological constant problem. Such a person needs to see what the 
essential point of the present paper is before committing to reading every 
secondary development. The essential point is obtained by reading 
Sections 2, 3, 4, 5 and perhaps 11. Readers who are 
content after reading these sections 
will certainly want to read the rest of the paper, 
which is however more technically 
involved.

\section{Approach to Euclidean Quantum Gravity}

\subsection{Unboundedness of Euclidean Einstein action}

In the Euclideanized version of quantum gravity \cite{hawking}, the 
 partition functional is given by a sum over Riemannian metrics of positive 
weights,
\begin{equation}
\label{gravintegral}
{\cal Z} = \int {\cal D} g_{\mu \nu} ~ e^{- S_{grav}}.
\end{equation}
The first important problem we must face is not a UV one but an IR one: 
the Euclidean 
Einstein action is unbounded from below \cite{hawking}, even when restricted 
to slowly varying geometries, 
and therefore the functional integral 
is ill-defined, even after UV regularization. 
Our approach to this deep problem is to avoid it and work 
completely perturbatively in the gravitational coupling and 
graviton field, $h_{\mu \nu} \equiv g_{\mu \nu} - \delta_{\mu \nu}$. 
Perturbatively, inverting 
the quadratic terms of $S_{Einstein}$ plus gauge fixing 
terms formally defines a Euclidean space propagator, while the remainder 
of the action plus ghost terms give Euclidean space interaction vertices. 
Thereby, one has a well-defined set of Feynman rules, despite the 
non-existence of the Euclidean path integral. The associated Feynman diagrams 
have meaning, they are the analytic continuation of the 
Feynman diagrams from Lorentzian spacetime. 
From now on we will consider Eq. (\ref{gravintegral}) to be just a formal 
mnemonic for the Euclidean quantum gravity perturbation expansion. At this 
level the only issue is the UV regularization.

\subsection{Choice of gravity cutoff}

We will find it convenient to switch language and denote the gravity 
cutoff by a distance scale $\ell$ rather than an energy scale. Mostly $\ell$ 
will be 
kept general, 
but later phenomenological considerations will constrain $\ell 
\sim {\cal O}(100)$ microns. 
A procedure is required  for coarse-graining the fundamental gravity sector
and matching it to effective gravity, valid above  distance scale $\ell$,
\begin{equation}
\label{gravl}
{\cal Z} = \int [{\cal D} g_{\mu \nu}]_{_{\ell}} ~ e^{- S_{grav}}.
\end{equation}
We will assume that there is a cutoff 
procedure (at least perturbatively about 
flat space), compatible with general covariance, which gives meaning to this 
expression. The detailed form is not required beyond knowing that it 
cuts off gravitational fluctuations below $\ell$.  

One concrete (but inelegant) example to keep in mind is just a  momentum 
cutoff, corresponding to
integrating out the high momentum modes of the 
gravitational field, $h_{\mu \nu} \equiv g_{\mu \nu} - \delta_{\mu \nu}$, 
so that the remaining measure 
$\int [{\cal D} g_{\mu \nu}]_{_{\ell}}$ has a momentum cutoff 
$|p| < 1/\ell$. While this preserves Euclidean ``Poincare'' invariance, 
it violates general coordinate invariance, or more precisely, after 
gauge fixing and adding ghost fields
 in some relativistic gauge such as de Donder, it violates 
BRST invariance. 
It is well 
known in quantum field theory that it is permissible to use a gauge
symmetry violating regulator as long as one adds counter-terms to the 
regulated theory so that the BRST identities are recovered in the 
continuum limit \cite{thooft} \cite{bns}. Here
we are not 
trying to take the UV cutoff away as in a real continuum limit, but 
rather ensure that the BRST identities are recovered for soft 
gravitational momenta, $|p| \ll 1/\ell$. This is done by adding and tuning 
BRST-violating counter-terms, order by order in gravitational momenta$\times 
\ell$, which restore the BRST symmetry of the soft amplitudes. This is 
analogous to restoring BRST symmetry in theories with a continuum limit, 
but with the technical difference that  in the renormalizable case the number 
of counter-terms is fixed, while in the present case the number of 
counter-terms increases with the order to which one is working in the soft 
momentum expansion. In particular, the procedure breaks down 
for gravitational momentum transfers $\sim 1/\ell$. This will not matter for us 
since BRST symmetry is important for (the Euclidean reflection of) unitarity, 
and we are already anticipating that it will be impossible to maintain 
exact unitarity for gravitational momentum transfers $\sim 1/\ell$. 
The basic procedure of symmetry restoration systematically 
in the soft momentum expansion was discussed in Ref. \cite{symanzik}.

The counter-terms to restrore BRST invariance in the infrared may
appear as a type of ``fine-tuning'', which  adds to and 
complicates the 
cosmological constant fine-tuning problem we are trying to solve. However, 
there is a clear distinction. The BRST-restoring counter-terms are added 
according to a symmetry principle, which we believe is a principle of 
the fundamental theory, but mutilated by the coarse-graining procedure 
whereby some degrees of freedom are integrated out and some are left in 
the theory.
These counter-terms are not a cheat when it comes to the cosmological 
constant, which is BRST-invariant and cannot be cancelled by the 
{\it minimal} set of BRST-restoring 
counter-terms.

There may of course be other more elegant cutoff procedures that we could use.
Dimensional regularization is certainly an elegant and simple gravity 
regularization. For most of this paper we will not consider it, partly 
because I have not yet checked its applicability to the type of 
integrals arising from the novel way that matter is treated here. It will 
however be very useful in phrasing the simple 
 prescription of Section 10, which 
gives a practical recipe for realistic computations while staying true to the
results of the previous sections.

\section{First Quantized Scalar Matter}

\subsection{Second-quantized ``target''}

Now let us add some matter to our theory. 
In this section we consider a ``target'' Euclidean field theory,
\begin{equation}
\label{target}
{\cal Z} =
 \int [{\cal D} g_{\mu \nu}]_{_{\ell}} {\cal D} \psi 
~e^{- S},
\end{equation}
where  the matter fields, $\psi_n$, are all scalars. This will keep the 
first quantized form as simple as possible. They may have any masses, heavy 
or light, but of course light scalars will require fine-tuning. 
This is an entirely separate fine-tuning from the cosmological 
constant problem and I will assume that the scalar fine tuning is performed 
to obtain any desired spectrum of physical masses. Scalars are just a 
simplifying step in the analysis.

While gravity wavelengths are somehow cut off at $\ell$,
 matter is of course cut off at much smaller 
distance. However by choosing a renormalizable form for the 
matter we can make the theory insensitive to the matter cutoff. In 
particular we will take an action of the form,
\begin{equation}
\label{psiS}
S = \int d^4x \sqrt{g} \{\frac{1}{2}  g^{\mu \nu} \partial_{\mu} \psi_n
\partial_{\nu} \psi_n + V(\psi) + K(\psi) R 
+ c R^2 + d R_{\mu \nu} R_{\mu \nu} \} + {\rm sources},
\end{equation}
where the potential $V$ is a quartic polynomial in the $\psi_n$, and 
$K(\psi)$ is quadratic in the $\psi_n$, with $K(0) 
\approx M_{Pl}^2$. We will allow $V$ to be rather general except that 
we assume that super-renormalizable interactions such as 
$\psi$ tadpoles and cubic potentials are weak enough relative to the 
relevant masses that their effects can be treated perturbatively. Also 
we assume that the field basis of the $\psi_n$ has been chosen so as to 
diagonalize their mass-matrix, and that we are expanding about 
an origin in $\psi$-space such that the mass-squared eigenvalues 
are all positive. 
By power-counting, the  couplings of Eq. (\ref{psiS}) are sufficient to 
renormalize the matter propagating in {\it fixed} geometries. The geometries 
themselves are being integrated over but with cutoff $\ell$. Thus 
our scalar matter analog of Eq. (\ref{2cutoff}) is well-defined 
(in perturbation theory). The higher derivative gravity couplings $c$ and 
$d$ (needed to renormalize matter loop divergences) are taken to be 
small enough to treated as perturbations as well.


\subsection{First quantized translation}

Now let us convert the perturbative expansion for Eqs. (\ref{target}) 
and (\ref{psiS}) to 
a first quantized form (for scalar matter). The simplest guess  for how
 to do this is
to employ path integrals over particle worldlines of mass $m$ using the 
(Euclideanized) point-particle action,
\begin{equation}
\label{1stquantized}
\int {\cal D} X(\tau) ~e^{- m s[X]},
\end{equation}
where $X^{\mu}(\tau)$ is the parametrized worldline and 
$s$ is its proper length  as measured 
by $g_{\mu \nu}$. This obviously has the right classical limit given by 
geodesics. However, the complicated square-root arising from Pythagoras' 
Theorem in infinitesimal contributions to $s$, and the 
inapplicability to  massless matter, $m = 0$, make this type of path 
integral problematic. These problems are resolved when an auxiliary 
worldline ``einbein'' field is introduced, giving more tractable 
path integrals of the form
\begin{eqnarray}
\label{ein}
 \int {\cal D} X^{\mu}(\tau) \int {\cal D} e(\tau)~
e^{- \int_0^1 d \tau \{g_{\mu \nu}(X) \dot{X}^{\mu} 
\dot{X}^{\nu}/2e
+ m^2 e/2 \}}.
\end{eqnarray}
It is understood that the reparametrization invariance of these path integrals
is to be treated as a gauge symmetry in the measure, so the measure has 
the volume of the associated gauge orbits divided out. The worldlines in 
such path integrals can either be taken as having fixed end-points in 
spacetime, thereby defining a scalar propagator in the background geometry, 
or as closed loops, thereby yielding a one-loop effective action for the 
background gravitational field. As long as the measure is generally 
coordinate invariant, the general 
covariance of the path integrals is manifest. 

Considerable effort has gone into taming path integrals of the form of 
Eq. (\ref{ein}), with special attention to the  
 measure, in order to demonstrate that 
they reproduce the standard second-quantized expectations up to 
renormalizations of matter and gravity couplings in Eq. (\ref{psiS}). 
Refs. \cite{polyakov} and \cite{joe} discuss the first-quantized approach 
in flat background spacetime, Ref. \cite{schubert}
reviews work on the first-quantized approach in non-trivial background 
geometry, and Ref. \cite{bastianelli} describes some recent progress. 
I will take it for granted 
here that Eq. (\ref{ein}) is well-defined and agrees with standard
 second quantization, at least for perturbative gravitational fields.
We will not need a detailed form of the measure. 
We will however 
need to maintain a compact notation, and for this reason (only) 
  {\it 
I will from now on write path integrals 
of the type in Eq. (\ref{1stquantized}) as a mnemonic for path integrals of 
the type in Eq. (\ref{ein}).} There is little danger of confusion since the central results we will need are some 
bounds on the Boltzman weights for non-zero $m$, 
which apply equally to either action. These bounds all follow from the 
fact that  for any $X(\tau)$, the Boltzman weight of 
Eq. (\ref{ein}) is at most as big as the Boltzman weight of 
Eq. (\ref{1stquantized}),  easily verified by first extremizing
 with respect to $e(\tau)$.

The other ingredients in the perturbative expansion for matter are the 
vertices connecting propagators. These can be read off from 
Eq. (\ref{psiS}). Note that there are no derivative couplings (in matter) 
so the couplings associated with the allowed vertices are just numbers. 
I will call a ``web'', $W$, a network of matter particle world-lines in 
Riemannian spacetime which 
end only at vertices following from the action Eq. (\ref{psiS}), or at 
sources. We thereby arrive at the first quantized form of the 
perturbative expansion of Eq. (\ref{target}),
\begin{equation}
\label{1quant}
{\cal Z} = \int [{\cal D} g_{\mu \nu}]_{_{\ell}} ~e^{- S_{grav}}
\int {\cal D} W 
e^{- \sum_n m_n s_n[W]} \times \lambda[W],
\end{equation}
where now
\begin{equation}
\label{gravityaction}
S_{grav} = \int d^4x \sqrt{g} \{ \Lambda_{cos} + M_{Pl}^2 R 
+ c R^2 + d R_{\mu \nu} R_{\mu \nu} \} + {\rm gravitational~ source},
\end{equation}
$s_n[W]$ is the length of the part of the web made out of world-lines of 
particle species $n$, and $\lambda[W]$ is the product of all the coupling 
constants at the vertices of $W$ that follow from Eq. (\ref{psiS}).

To summarize, Eq. (\ref{target}) is an analog of the real world effective 
description, Eq. (\ref{2cutoff}), but with perturbative scalar matter, 
and Eq. (\ref{1quant}) is its first-quantized form, with gravity still being
treated as a (coarse-grained) field. As claimed in the introduction, 
Eq. (\ref{target}) is really not a sufficiently 
minimal description. 
It includes a strong prejudice about the short-distance 
coupling of gravity to matter. But to see this we must massage 
Eq. (\ref{1quant}). 

Eq. (\ref{1quant}) can straightforwardly be rewritten,
\begin{equation}
\label{connected}
{\cal Z} = \int [{\cal D} g_{\mu \nu}]_{_{\ell}} ~e^{- S_{grav}}~
e^{\int {\cal D} w 
e^{- \sum_n m_n s_n[w]} \times \lambda[w]},
\end{equation}
where 
$w$ denotes connected webs (where ``connected'' 
means only using matter lines, 
not gravity lines), while $W$ denotes a general web (that is,
some collection of connected webs). The exponentiation occurs for the 
same reason as in Feynman diagrams, because in the original sum, when there 
are $N$ connected components we have a factor $1/N$! in order to identify 
the different possible permutations.

\section{Minimal effective description and Fat Gravity model}

We can decompose the path integral over
connected webs into a sum over two classes, say $A$ and $B$,
\begin{equation}
\int {\cal D} w = \int [{\cal D} w]_{A} + 
\int [{\cal D} w]_{B}.
\end{equation}
Here, class $A$ consists of any ``vacuum'' web that 
has ``diameter'' less than $\ell$. 
The precise definition of ``vacuum'' web is one where no worldlines connect 
to external sources, although they may end on perturbative 
tadpole couplings in the 
potential $V$ in Eq.(\ref{psiS}). This is very much the analog of second 
quantized vacuum diagrams. The quotation marks  only remind us 
that we are treating the gravitational field as an aspect
 of the vacuum in this definition. We define the diameter of a connected 
vacuum web  as the maximum geodesic distance (given the particular background 
metric $g_{\mu \nu}$) between any two points 
on the web (though the geodesic between these two points is generally off 
the web). Class B is simply the complement of class 
A. Intuitively, class A are ``small'' vacuum webs and class B are the 
remaining webs.

Thus,
\begin{eqnarray}
\label{product}
{\cal Z} &=& \int [{\cal D} g_{\mu \nu}]_{_{\ell}} ~e^{- S_{grav}} 
e^{\int [{\cal D} w]_{A} e^{- \sum_n m_n s_n[w]} \times \lambda[w]}
~ e^{\int [{\cal D} w']_{B} e^{- \sum_n m_n s_n[w']} \times \lambda[w']} 
\nonumber \\
&=& \int [{\cal D} g_{\mu \nu}]_{_{\ell}} ~e^{- S_{grav}} 
\int [{\cal D} W]_{A} e^{- \sum_n m_n s_n[W]} \times \lambda[W] \nonumber \\
&~& ~ ~ ~ ~ ~ ~ ~ ~ ~ ~ \times
\int [{\cal D} W']_{B} e^{- \sum_n m_n s_n[W']} \times \lambda[W'],
\end{eqnarray}
where in the second equality we are going back to the sum over all webs 
without regard to connectedness. For the second equality to hold we 
must generalize the definition of classes A and B to webs with disconnected 
components. The generalized definition is (obviously) 
that class A consists of only 
vacuum webs, each connected component of which has diameter less than $\ell$. 
Class B consists of webs where each connected 
component is either a non-vacuum web, or a vacuum web with diameter greater 
than $\ell$.

The great virtue of this way, Eq. (\ref{product}),
 of presenting the first quantized 
form of Eq. (\ref{target}), is that it is now manifest that there was 
in fact some more coarse-graining that we should have done in order to get 
a truly minimal effective description. We have written the partition 
functional as a product of path integrals. 
Coarse-graining means doing some of the 
integrals over physics that is not experimentally accessible and leaving the
 remaining integrals. In second quantized form, Eq. (\ref{2cutoff}) or the
scalar analog Eq. (\ref{target}), it is not obvious that there is any more 
coarse-graining we could do without integrating out tested physics, but 
in Eq. (\ref{product}) we clearly see that the integral over webs of class
A can still be done. Since they are all vacuum webs they do not contribute 
to {\it any} amplitudes with matter external lines, but only to the 
pure gravitational effective action. In general such contributions to the 
gravitational effective action would be 
non-local, but   since all the class A  webs are 
``small'', their effect on 
the coarse-grained gravitational 
fields (without fluctuations smaller than $\ell$) can be matched 
by purely local terms in the effective action. Furthermore, the 
decomposition of webs into classes A and B is obviously generally covariant, 
so the dominant terms in the local effective action arising from integrating 
out class A is the same as the terms we have already included in the 
gravitational action, Eq. (\ref{gravityaction}). That is, {\it 
integrating out 
class A should be considered as part of the 
process of gravitational coarse-graining up to distance} $\ell$. 

After integrating out class A, our minimal effective description is 
\begin{eqnarray}
\label{Balone}
{\cal Z} &=&  \int [{\cal D} g_{\mu \nu}]_{_{\ell}} e^{- S_{grav}} 
\int [{\cal D} W]_{B} ~ e^{- \sum_n m_n s_n[W]} \times \lambda[W].
\end{eqnarray}
Note that this description has the same power as the less minimal Eqs. 
(\ref{2cutoff}) and (\ref{target}). All non-gravitational matter amplitudes 
arise from class B webs and are unchanged from Eq. (\ref{target}) or its 
first-quantized equivalent, Eq. (\ref{product}). Matter couplings to 
a fixed background metric are also unchanged and obey the Equivalence 
Principle, again because we have not touched the non-vacuum webs. 
For gravitational momentum transfers far below $1/\ell$, we have the usual 
local and general coordinate invariant effective action, locality after 
integrating out class A being argued in the previous paragraph. 
The purely matter amplitudes match the analytic continutations of the usual 
unitary Lorentzian ones, as do the full amplitudes with soft 
gravitational momentum transfers $\ll 1/\ell$. In this way, our
checklist from the introduction, (i -- vii), is satisfied, if we 
make an appropriate allowance for scalar matter in interpreting (i).

Note that the soft gravitational 
amplitudes are not at all trivial. For example, 
they contain standard non-analyticities arising from ``large'' 
diameter class B vacuum loops of
 very light and soft scalar matter (the analog of soft photons say). 
That is they match the 
analytic continuation from 
Lorentzian spacetime of soft gravity amplitudes 
with imaginary parts corresponding to intermediate 
processes of the form,
\begin{equation}
{\rm soft~ gravitons} \rightarrow {\rm soft~ matter}.
\end{equation}
The integrated out class A vacuum loops do not affect these non-analyticities.
As argued above they only renormalize (analytic)
 gravity effective vertices for coarse-grained gravity. 

\subsection{Technically natural size of the cosmological constant}

Of course, the important question is what is the size of the cosmological 
constant in Eq. (\ref{Balone}) now that we have integrated out the class A 
webs. If we had started from (the scalar-matter analog of) Eq. (\ref{teveft}),
then the class A webs would contribute a large renormalization of the 
cosmological constant.
But the central point of the present paper is that without committing to 
a particular model of short-distance ($< \ell$) gravity-matter coupling 
we do not know the contribution of class A webs. They are simply not part of 
our minimal effective description.
 The most conservative thing we can do is to  estimate the 
renormalization-stable size {\it within} this description, 
Eq. (\ref{Balone}). For simplicity, 
let us do this for the case where any matter mass eigenvalue is
 either ``heavy'' or ``light'', $m_n \gg 1/\ell$ or $m_n \ll 1/\ell$. 
There are two types of loops to consider when estimating the renormalization 
stable value of the cosmological constant, matter loops and loops involving 
graviton exchange. Let us first consider pure matter loops, arising from 
summing vacuum webs in class B with fixed topology (where particle-type and 
vertex-type are considered as features of the topology of the web). 
If a web consists of purely heavy matter, then it makes a negligible 
contribution to the infrared cosmological constant, suppressed by at least 
$e^{-m_n \ell}$, since $\ell$ is the minimal vacuum-web diameter in class B. 
Thus, these heavy matter contributions to the cosmological constant which are 
robustly large in the effective descriptions of Eqs. (\ref{2cutoff}) and 
(\ref{target}), 
are negligible in the effective description of Eq. (\ref{Balone})!

The only 
webs that contribute to the gravitational effective action 
must come from light matter straddling distances 
of $\ell$ or larger, with any heavy matter propagating over short 
distances. Such short lines can be treated as approximate local effective 
vertices from the perspective of the light matter propagating over $\ell$ 
distances. That is, for this class of webs we could get the same answer by 
having first integrated out all heavy masses at the very beginning of our 
story, Eq. (\ref{target}), so that their sole effect is to renormalize the 
vertices for light matter. Then we can just focus on light matter loops as if 
there were no heavy particles.  One can imagine 
light matter diagrams 
regulated by heavy Pauli-Villars fields, in which case we immediately 
see that these  regulators decouple from the gravitational effective action
 for the same reason as 
heavy physical particles do, other than a renormalization of light matter 
vertices. Thus after light matter renormalization, 
 class B contributions to the gravitational 
effective action must be UV cut off by $1/\ell$. There is no other 
scale since we 
can neglect the light masses. The light 
matter contribution to the cosmological 
constant must therefore be of order $1/\ell^4$.
Of course, graviton momenta, and therefore their 
corrections to the cosmological constant,
 are also cut off by $1/\ell$. 

In summary, the technically natural size of the 
cosmological constant in the effective description of Eq.(\ref{Balone}) is 
$\sim 1/\ell^4$. This completes step I described in the introduction for the 
simple case of perturbative scalar matter.

\subsection{Euclidean Fat Gravity Model}

Although we know the technically natural size of the cosmological constant,
arising from the physics described by Eq.(\ref{Balone}),  
we do not know the contributions from the physics integrated out, namely 
the gravitational sector at short distances $< \ell$ and its couplings 
to ``small'' class A matter webs. It is now easy to give a toy 
model of what this physics might look like which has the defining features 
of what I have previously called ``fat gravity''. We will take the 
gravitational sector to simply not allow fluctuations below distances 
of $\ell$, much as perturbative strings cannot have meaningful fluctuations 
below $\ell_{string}$.  Our coarse-graining cutoff provides a crude model of 
this. Secondly, we assume that fat gravity is blind to the small 
class A matter webs. That is, the renormalization of the 
gravitational effective action from integrating out class A webs is 
negligible. Finally, we will take the tree-level cosmological constant, 
$\Lambda_{cos}$, to vanish or be small $< 1/{\ell}^4$.

With these assumptions, our toy model of fat Euclidean gravity is simply 
to take Eq. (\ref{Balone}), not as an effective description, but as the 
full model! Clearly, in the toy model then, our estimate of the 
technically natural size of the cosmological constant is its real size.
Another way to write the fat gravity model  is 
\begin{eqnarray}
\label{model1}
{\cal Z}_{fat~gravity} 
&=&  \int [{\cal D} g_{\mu \nu}]_{_{\ell}} e^{- S_{grav}} 
\int {\cal D} W ~ e^{- \sum_n m_n s_n[W]} \times \lambda[W] \times
 \theta_B[W],
\end{eqnarray}
where now we are formally integrating over {\it all} webs, but 
the usual weight for a web is multiplied 
by $\theta_B[W]$, which is one if the web is in class B, and zero 
if the web is in class A. 
This rather trivial re-writing makes more manifest that the 
short-distance modification of gravity coupling to matter, represented by 
$\theta_B[W]$, has absolutely no dependence on
 the matter couplings and masses, 
which appear in the other factors in the weight of a web. There is no 
secret fine-tuning with respect to the matter parameters. Yet it suppresses 
the cosmological constant corrections. That is remarkable from the usual 
viewpoint. Furthermore, $\theta_B[W]$ is completely generally coordinate 
invariant. It is not absolutely local as a functional of the web, but 
it is local down to the distance $\ell$. This is possible in a fat graviton 
model, the graviton should really be viewed as spread 
over a distance $\ell$. 

Approaches to the cosmological constant problem which
take non-locality as an underlying  physical principle are 
Ref. \cite{moffat} and Refs. \cite{savas}. 
Ref. \cite{moffat} entertains non-locality
only up to distances of order 100 microns, and in this sense crudely
resembles 
the approach of this paper, without however having safe-guarded 
the Equivalence Principle for quantum matter. The non-locality in Refs. 
\cite{savas} on the other hand is present  at the largest distances.
This approach seems quite orthogonal to that pursued here.



Our toy model has sharp transitions at distance $\ell$, which 
in any fully realistic relativistic model would be smoothed out and 
 correlated with gravitational 
excitations at $1/\ell$. These excitations are 
 obviously missing in our toy model and mean that for  precisely 
momentum transfers of order $1/\ell$ it has no 
unitary analytic continuation to Lorentzian spacetime. For much 
higher momentum transfers, in our model only pure matter 
interactions are possible, and these are obviously the continuation of 
unitary matter interactions from Lorentzian spacetime since we have 
 not tampered with 
the non-vacuum webs. Also for very small momentum transfers $\ll 1/\ell$
the model amplitudes (now both gravitational and light matter)
 are continuations of unitary Lorentzian amplitudes,  
just as when Eq. (\ref{Balone}) 
is read as a coarse-grained effective description.

This completes step II of the program discussed in the introduction, for the 
simplest type of perturbative scalar matter.

\section{Phenomenology of Fat Gravity}

A rather straightforward consequence of fat gravity, Eq. (\ref{model1}), 
is that the static gravitational force, 
which is Newtonian at distances $\gg \ell$, must be 
 suppressed at short distances 
$< \ell$. The usual 
long distance behavior follows because we are not modifiying 
gravity in that regime. In momentum space the static (zero energy) behavior
is just given by the graviton propagator $\sim 1/\vec{q}^2$. 
At short distances $< \ell$ however, 
the gravitational field is cut off and the 
static force is suppressed. This is what makes the fat gravity model 
phenomenologically interesting, it has this rather unique
 testable prediction. The 
suppression of the gravitational force at short distance 
in effective field theory can only be accomplished by adding short 
distance repulsive forces of equal strength. This equality of force strengths is 
technically unnatural in non-supersymmetric effective field theory. Thus 
seeing a significant suppression would indicate non-field-theoretic 
gravitational physics, such as fat gravity.

One might wonder whether we really had to impose the gravity cutoff at $\ell$,
thereby modifying short-distance static gravity. If we had not done this and 
continued the Newtonian force to short distances and high momenta,
 we could  
rotate such graviton exchanges using $SO(4)$-symmetry  into the  
Euclidean 
continuation of hard timelike graviton exchanges. Unitarity would then 
require non-vanishing gravitational effective action sensitive to hard/heavy
 matter loops, with Lorentzian continuation having imaginary parts 
corresponding to 
\begin{equation}
{\rm timelike ~graviton} \rightarrow {\rm hard/heavy ~matter}.
\end{equation}
But we have shown that such sensitivity is absent in  fat gravity. 
Consistency of the
fat graviton model therefore 
 requires that the gravitational coupling to matter
 be suppressed at short distance. 

There is a very narrow window for fat gravity to solve the 
cosmological constant problem. Gravity is tested down to $\sim 100$ microns 
\cite{gravexpt1} 
without deviation from Newtonian predictions, so $\ell < 100$ 
microns. On the other hand, the natural size of the 
radiatively stable cosmological constant in fat gravity is $\sim 
1/(16 \pi^2 \ell^4)$, to be compared with the observed dark 
energy $\sim (10^{-3}$ eV$)^4$ \cite{ccexpt} \cite{pdg}. 
Therefore, naturalness imposes $\ell > 20$ 
microns \cite{old}.

\section{``Block-Spin'' coarse-graining: all-scalar warm-up}

\subsection{Gravity $\rightarrow$ scalar model}

We will now pursue a technically different coarse-graining procedure 
which will allow us to generalize our first quantized observations to 
standard second quantization and general matter content, 
thereby allowing us to tackle realistic and non-perturbative 
matter phenomena, including issues such as multiple vacua. 
We proceed in two stages. As a warm-up, but not strictly 
necessary, we look at a theory without any type of gauge symmetry, 
involving two sectors both containing only scalars. That is, we not only 
consider matter to be scalars, $\psi$, but replace the gravity sector
by a single light real scalar $\phi$. The metric in the model is now fixed 
to flat space.
Our starting point in 
second-quantized language is then
\begin{equation}
\label{targetphi}
{\cal Z} = \int [{\cal D} \phi]_{_{\ell}} {\cal D} \psi 
e^{- S}, 
\end{equation}
where the action has scalar interactions, and 
 the $\phi$ sector has been cut off at $\ell$ in 
the spirit of coarse-graining. For example, the
 $\phi$ cutoff can be taken as a momentum 
cutoff $p^2 < 1/\ell$. Unlike the case of real gravity discussed in Section 
2.2, this cutoff violates no essential 
gauge or global symmetries.
We imagine that $\phi$ couplings to $\psi$ 
matter and to itself 
are extremely weak and observable only at small momentum transfers 
compared to $1/\ell$ for similar reasons to the real gravity case. 
To parallel the usual gravity notation we will take $\phi$ to be dimensionless,
with a large ``Planck scale'' normalization, 
suppressing its interactions (with 
other dimensionful scales set by TeV or less).
For a technical reason, not present for realistic matter, we assume that 
there is an exact symmetry $\psi \rightarrow - \psi$ for  {\it light} matter 
fields.

 We can now write the effective description Eq. (\ref{targetphi}) 
with $\psi$ matter in first quantized form,
\begin{eqnarray}
\label{productphi}
{\cal Z} &=& \int [{\cal D} \phi]_{_{\ell}} e^{- S_{\phi}} ~
e^{\int [{\cal D} w]_{C} ~e^{- \sum_n m_n s_n[w]} \times \lambda[w]}
~ e^{\int [{\cal D} w']_{D} ~e^{- \sum_n m_n s_n[w']} \times \lambda[w']} 
\nonumber \\
&=& \int [{\cal D} \phi]_{_{\ell}} e^{- S_{\phi}} 
\int [{\cal D} W]_{C} ~e^{- \sum_n m_n s_n[W]} \times \lambda[W]
\int [{\cal D} W']_{D} ~e^{- \sum_n m_n s_n[W']} \times \lambda[W'],
\end{eqnarray}
where we make a decomposition of webs into two classes 
C and D, which is technically different from our earlier decomposition into 
classes A and B, but morally the same. 
Connected vacuum webs (where now ``vacuum'' includes the $\phi$ background 
just as it did the gravitational background earlier) 
 with diameters $\ll \ell$ are all in C, while 
connected non-vacuum webs and 
connected vacuum webs with diameters $\gg \ell$ are all in D, just as 
was the case before for A and B respectively. The detailed 
differences concern 
vacuum webs with diameters $\sim {\cal O}(\ell)$.

The C/D decomposition is obtained by first imposing a hypercubic lattice 
structure on the Euclidean spacetime, with lattice spacing of $\ell$, and 
elementary blocks of size $\ell^4$. These blocks will provide a fixed 
basis for distinguishing large and small webs, rather than using 
web diameters. Although the diameter approach is certainly the more 
elegant and spacetime-symmetric one, the  approach we now take, 
related to standard ``block-spin'' coarse-graining \cite{block-spin},
 makes the exercise in 
accounting we are about to do more tractable. 
Since the precise definitions of C and D, and the statement and 
proofs of their properties, is rather technical, I will first give a 
sloppy description to get across the basic idea. 
``Small'' class C webs are typically 
contained in some 
particular hypercubic block, denoted by integer coordinates, $N^{\mu}$. 
Thus, 
\begin{equation}
\label{slop}
\int [{\cal D} w]_{C}  ~ e^{- \sum_n m_n s_n[w]} \times \lambda[w]
\approx \sum_{N} \int [{\cal D} w]_{N} ~
e^{- \sum_n m_n s_n[w]} \times \lambda[w],
\end{equation}
where the integral on the right-hand side is over all connected vacuum webs 
which are entirely within block $N$. Now such an integral 
over webs with restricted 
domain has a simple and obvious second quantized form,
\begin{equation}
\label{2q}
\int [{\cal D} w]_{N} ~  e^{- \sum_n m_n s_n[w]} \times \lambda[w]
 = \ln \{ \int_{\psi(\partial N) =0}  
 {\cal D} \psi e^{-S_{matter}} \},
\end{equation}
where the integral on the right hand side is over fields inside block $N$ 
with Dirichlet boundary conditions on the boundary of the block, and 
$S_{matter}$ is the part of the action depending on matter fields (and 
possibly $\phi$ too). The 
logarithm arises because we have restricted to connected webs. Thus we 
can write a fully second quantized version of our ultimate coarse graining 
procedure (I) and the associated ``fat $\phi$'' model (II). 
The only problem is that 
a small but non-negligible fraction of small webs 
will not sit entirely within some hypercubic block, but will straddle a 
boundary between two blocks, hence the ``$\approx$'' in Eq. (\ref{slop}). 
 We will be much more careful below.

\subsection{Pure mathematics of blocks and webs}

The results we need are given here and proven in the appendix. 
There exists a subset of 
connected vacuum webs, C, such that any integral over C (with arbitrary 
integrand) can be re-written,
\begin{equation}
\label{Csum}
\int [{\cal D} w]_{C} ... = \sum_{N} \sum_{b \in {\cal S}} \epsilon(b) 
\int [{\cal D} w]_{b_N} ...,
\end{equation}
where C contains every connected vacuum web with diameter $< \ell$, 
some with $\ell \leq$ diameter $\leq 4 \ell$, and none with 
diameter $> 4 \ell$ (that is, C defines a precise notion  of ``small'' 
vacuum webs). The set ${\cal S}$ is a finite set of ``composite blocks'', $b$, 
all contained in the region of spacetime $[0, 2 \ell]^4$. Composite blocks 
are simply unions of some of the elementary $\ell^4$ blocks of the
hyper-cubic lattice. $\epsilon(b) = \pm 1$ depending in some way on the 
block $b$. The composite block $b_N$ is simply the block $b$ translated 
by $N^{\mu} \ell$ from the vicinity of the origin to the vicinity of the 
elementary block $N$. 
The integral on the right hand side is over all connected vacuum webs 
entirely contained inside block $b_N$.

A corollory of the generality of Eq. (\ref{Csum}) for general 
integrands is that even though the 
right hand side sums over webs with varying signs and multiplicities 
(since the blocks $b \in {\cal S}$ can overlap), {\it in net} any web gets summed 
once or not at all. Those that are summed are in C and those 
that are not (or are non-vacuum connected webs) are defined to be in class D. 

In the case where the web integrand is the standard one, we can clearly
write a simple second-quantized translation analogous to Eq. (\ref{2q}), 
\begin{equation}
\label{2qb}
\int [{\cal D} w]_{b_N} e^{- \sum_n m_n s_n[w]} \times \lambda[w] 
= \ln \{ \int_{\psi(\partial b_N) =0} 
 {\cal D} \psi e^{-S_{matter}} \},
\end{equation}
where the integral on the right-hand side is over fields within $b_n$ 
which obey Dirichlet boundary conditions on the boundary of $b_N$.

Two more results we will need involve integrals over the insides and 
boundaries of our blocks, $b_N$,
\begin{equation}
\label{bulk}
\sum_{N^{\mu}} \sum_{b \in {\cal S}} \epsilon(b) \int_{b_N} d^4x ... = \int d^4x ...,
\end{equation}
where the right hand side is the integral over the entire Euclidean 
spacetime, and 
\begin{equation}
\label{bdry}
\sum_{N^{\mu}} \sum_{b \in {\cal S}} \epsilon(b) \int_{\partial b_N} d^3x ... = 0.
\end{equation}

The explicit construction of the composite blocks $b \in {\cal S}$ that ultimately 
define classes C and D and their properties is detailed in the appendix. 
However,
 other than their existence we will not need their explicit form here.

\subsection{The more minimal effective description}

By the above results we have the second quantized translation,
\begin{equation}
\label{2qC}
\int [{\cal D} w]_{C} ~ e^{- \sum_n m_n s_n[w]} \times \lambda[w]
= \sum_{N} \sum_{b \in {\cal S}} \epsilon(b) 
\ln \{ \int_{\psi(\partial b_N) =0} 
 {\cal D} \psi e^{-S_{matter}} \}.
\end{equation}
The minimal effective description upon integrating out class C is 
therefore given by 
\begin{eqnarray}
\label{target-C}
{\cal Z} 
&=& \int [{\cal D} \phi]_{_{\ell}} e^{- S_{\phi}} 
\int [{\cal D} W]_{D}~  e^{- \sum_n m_n s_n[W]} \times \lambda[W] \nonumber \\
&=& \int [{\cal D} \phi]_{_{\ell}} e^{- S_{\phi} -
\sum_{N} \sum_{b \in {\cal S}} \epsilon(b) 
\ln \{ \int_{\psi'(\partial b_N) =0} 
 {\cal D} \psi' e^{-S_{matter}} \} }  
\int 
 {\cal D} \psi e^{-S_{matter}}.
\end{eqnarray}
In writing the second-quantized form in the second line we have used the fact 
that the connected webs of class D are all connected 
webs {\it minus} those in 
class C.
That is relative to Eq. (\ref{targetphi}), there are a set of 
``counterterms'' in the $\phi$ action coming from the absence of 
class C. 

These ``counterterms'' are manifestly local down to $\sim \ell$ in $\phi$, 
that is a sum over the lattice of terms which depends on  $\phi$ in regions 
of size of order $\ell^4$. These counterterms are not independent of 
matter couplings and masses, so from the second quantized viewpoint it 
looks as if we are including some finely tuned counter-terms, but in fact 
we have derived them from a first-quantized theory where we are simply 
integrating out short-distance gravity-matter 
physics about which we are ignorant (step I of our program). 
As before, we can consider 
Eq. (\ref{target-C}) as a toy model of a ``fat $\phi$'' (step II of our 
program), but this does not affect the mathematics.

Since the $\phi$ fields are cut off to only fluctuate on distances $> \ell$ 
these counter-terms can be matched to exactly local ones, that is, just 
renormalizing the couplings of $S_{\phi}$. Of course the class C 
counter-terms only respect hypercubic lattice symmetries, but full 
Euclidean spacetime ``Poincare'' symmetry is an automatic accidental 
symmetry of the leading operators in $\phi$. This is a standard infrared 
feature of many lattice theories. We will not be so fortunate
 in the case of real 
gravity.


Note that since the only difference between Eq. (\ref{target-C}) and 
Eq. (\ref{targetphi}) 
are $\phi$ counterterms which are local (down to $\ell$), 
Eq. (\ref{target-C}) has precisely the same non-analyticities in the 
low-energy, $\ll 1/\ell$, amplitudes, including those involving $\phi$. These 
are the Euclidean reflection of the imaginary parts of the Lorentzian 
diagrams corresponding to very low energy unitarity. If we go through the 
checklist (i -- vii) given in the introduction we see that we have now lost 
exact (Euclidean) Poincare invariance because of our cutoff. But the 
 symmetry is 
respected in the absence of the $\phi$ sector and is also recovered 
accidentally in the soft $\phi$ limit. There is no Equivalence Principle or 
real world data to test for. Other than these necessary exceptions other 
properties clearly hold.

\subsection{Suppression of $\psi$ contributions to $\Gamma_{eff}[\phi]$}

To quickly see the power of Eq. (\ref{target-C}), 
let us first specialize to the 
case where all matter is heavy, $m_n \gg 1/\ell$. Therefore  we can 
integrate it out completely in the integrals on 
 the right hand side of Eq.(\ref{2qC}) to yield an effective 
lagrangian for $\phi$ which is local. 
This local lagrangian can be divided up into bulk terms in each $b_N$ 
as well as terms localized on the boundary of the block, $\partial b_N$,
\begin{equation}
\int [{\cal D} w]_{C} ~e^{- \sum_n m_n s_n[w]} \times \lambda[w] = 
- \sum_{N} \sum_{b \in {\cal S}} \epsilon(b) 
 \{ \int_{b_N} d^4x {\cal L}_{bulk}(\phi)
+ \int_{\partial b_N} d^3x {\cal L}_{boundary}(\phi) \}.
\end{equation}
Note that since the lagrangians are local, the bulk lagrangian has 
no dependence on the nature of the block $b_N$. That is 
${\cal L}_{bulk}(\phi)$ is just the standard result in infinite Euclidean 
spacetime from integrating out the heavy matter. 

By Eqs. (\ref{bdry}) and (\ref{bulk}), 
the sum of all boundary terms cancel and the sum over bulk terms gives an 
integral over all Euclidean spacetime, 
\begin{equation}
\int [{\cal D} w]_{C} ~e^{- \sum_n m_n s_n[w]} \times \lambda[w] = 
 - \int d^4x {\cal L}_{bulk}(\phi).
\end{equation}
Thus class C contains precisely the corrections to the $\phi$ sector arising 
from integrating out matter in standard effective field theory. It follows 
that to all orders in the effective field theory expansion in 
$1/(m_{\psi} \ell)$, the class D  contribution 
to the $\phi$ effective action vanishes! This agrees with our old 
first-quantized argument: since all connected vacuum webs in D have 
diameter $> \ell$ and heavy particle lines, their contributions are 
necessarily suppressed by at least $e^{-m_{\psi} \ell}$. Thus the 
minimal effective description, Eq. (\ref{target-C}), 
does not suffer from a fine-tuning problem. Only coarse-grained 
$\phi$ loops can contribute to the $\phi$ potential, but this is all 
cut off  at $1/\ell$. 

Now let us consider a more general situation where 
some of the $\psi_n$ are very light, $m_n \ll 1/\ell$ and the rest are 
heavy, $m_n \gg 1/\ell$. To compute the right-hand side of Eq.(\ref{2qC}) 
we first integrate out all the heavy $\psi$'s, leaving a local 
effective lagrangian in each block $b_N$, for the $\phi$ and the light 
$\psi$'s. The story of the local terms which are independent of the light 
$\psi$'s is identical to the case above where these fields were absent. 
The boundary terms dependent on the light $\psi$'s must have derivatives 
into the bulk acting on them because the $\psi$'s themselves vanish at 
the boundary. Using this and the $\psi \rightarrow - \psi$ symmetry, 
by power-counting the only local lagrangian terms which can depend on 
positive powers of $m_{heavy}$ are the bulk light $\psi$ mass terms. 
However, these are precisely the terms which we are fine-tuning 
to ensure that our light scalar matter is in fact very light. We 
have accepted this purely matter fine tuning as the price for playing 
with scalar matter at all. Once done there are no terms in the bulk or 
boundary local lagrangians with couplings which  depend on
 positive powers of $m_{heavy}$, 
only ln$(m_{heavy})$  
or negative powers of $m_{heavy}$. 

Let us use these observations to estimate matter contributions to the 
$\phi$ effective potential in the minimal effective description, 
Eq. (\ref{target-C}).
Clearly, 
\begin{equation}
\label{effacphi}
\Gamma_{eff}[\phi] = \ln(\int 
 {\cal D} \psi e^{-S_{matter}}) - \sum_{N^{\mu}} \sum_{b \in {\cal S}} 
\epsilon(b) 
\ln \{ \int_{\psi(\partial b_N) =0} 
 {\cal D} \psi e^{-S_{matter}} \}, 
\end{equation}
again since class D connected webs are all webs {\it minus} class C.
By the procedure of integrating out heavy matter and matching to a matter 
theory with only light $\psi$'s, we can take the functional integral to be
over only these light fields, with $S$ replaced by the matched effective
vertices discussed above. By our discussion above, the 
diagrams involving only heavy particle lines cancel out. Furthermore, the 
diagrams involving light $\psi$ lines are UV finite {\it after} matter 
renormalization. The reason is that 
any divergence is local, just as pure heavy loops are local. 
Bulk divergences from class D ($=$ all connected 
webs minus class C) then
cancel by Eq. (\ref{bulk}), and 
boundary localized divergences cancel by Eq. (\ref{bdry}). Finally, the 
finite effective potential for $\phi$
 is set by the scale $1/\ell$, when we neglect light masses, 
 and non-negative powers of 
$1/m_{heavy}$ as power-counted above. 
That is, the corrections are all at most of order $1/\ell^4$.
(Recall that we normalized $\phi$ to be dimensionless like the metric.)

\section{Block-spin coarse-graining with real gravity and scalar matter} 


If matter is kept restricted to scalar fields (with $\psi \rightarrow -\psi$ 
symmetry for light matter as before), but $\phi$ is now replaced by 
real gravity, we can re-do the  steps of the previous section, and must only 
deal with the extra steps arising from the explicit breaking  
of the gravitational BRST symmetry by the coarse-graining.

The minimal effective description parallel to Eq. 
(\ref{target-C}) is then, 
\begin{eqnarray}
\label{target-Cgrav}
&{\cal Z}& 
= \int [{\cal D} h_{\mu \nu} ~ {\cal D} 
{\rm ghosts}]_{_{ \ell}} ~e^{- S_{grav} 
- S_{c.t.}} 
\int [{\cal D} W]_{D} ~e^{- \sum_n m_n s_n[W]} \times \lambda[W] \nonumber \\
&=& \int [{\cal D} h_{\mu \nu} ~{\cal D}{\rm ghosts}]_{_{ \ell}} 
~e^{- S_{grav} - S_{c.t.} - 
\sum_{N} \sum_{b \in {\cal S}} \epsilon(b) 
\ln \{ \int_{\psi'(\partial b_N) =0} 
 {\cal D} \psi' e^{-S_{matter}} \} } \nonumber \\ 
&~& ~ ~ \times \int 
 {\cal D} \psi e^{-S_{matter}},
\end{eqnarray}
where $S_{c.t.}$ represents the BRST-violating counter-terms to be tuned 
to recover BRST symmetry in the limit of soft gravity processes, already 
discussed in 
Section 2.2 in the case where a momentum cutoff is used for gravity.
The only difference here is that the set of required counter-terms is
 larger, respecting only hypercubic lattice symmetries, but otherwise 
the procedure and philosophy is the same as described in Section 2.2.
A similar and well-known procedure for restoring BRST symmetry associated to 
chiral gauge invariance in lattice-regulated theory was described in 
Ref. \cite{rome}. (This procedure was only established perturbatively, 
and indeed there is some  
controversy on the non-perturbative applicability. Of course here we are only 
proceeding perturbatively in $G_N$.)
 Analogously to 
 the $\phi$ case it is manifest that the non-analyticity in
 amplitudes for momenta below 
$1/\ell$ are the same as in Eq. (\ref{target}), since the difference consists 
of gravitational vertices which are local down to $\ell$.

Let us consider the matter contributions to the gravitational effective 
action, the analog of Eq. (\ref{effacphi}),
\begin{equation}
\label{graveffS}
\Gamma_{eff}[h_{\mu \nu}] = \ln(\int 
 {\cal D} \psi e^{-S_{matter}}) - \sum_{N} \sum_{b \in {\cal S}} \epsilon(b) 
\ln \{ \int_{\psi(\partial b_N) =0} 
 {\cal D} \psi e^{-S_{matter}} \}. 
\end{equation}
Once again we can integrate out heavy matter, the only sensitive 
terms to this in the light effective theory being BRST-violating or the 
light mass terms, all of which we are content to fine-tune away. 
There are no truly divergent BRST-violations since they would be local, and 
by Eqs. (\ref{bdry}) and (\ref{bulk}) these would cancel out of the class D 
contributions. Therefore 
integrating out the light matter now gives contributions to the 
gravitational effective action which are set by the scale $1/\ell$, 
whether BRST conserving or not, this being the only scale since divergences 
cancel as usual using Eqs. (\ref{bulk}) and (\ref{bdry}), 
and light masses are 
neglected. For the cosmological constant the 
dominant contributions must be of order $1/\ell^4$ with corrections by 
powers of $1/(m_{heavy} \ell)$. BRST and continuous spacetime symmetry 
violations are suppressed in the 
infrared by tuning of $S_{c.t.}$.

When matter consists only of heavy fields, the separation 
of BRST violation and the cosmological constant problem is 
particularly clear. In this case 
the right hand sides of the matter contributions to the gravitational 
effective action, Eq. (\ref{graveffS}), after integrating out the heavy 
matter is given by integrals of local effective lagrangians 
(for $h_{\mu \nu}$), of either bulk or boundary type. 
By Eqs. (\ref{bulk}) and (\ref{bdry}) the entire matter contribution to 
the gravity effective action vanishes! In particular there is no 
correction to the cosmological constant and there is no violation of 
BRST symmetry from the matter. The only violation of BRST invariance is
at most from  
gravity loops due to a non-symmetric gravity regulator itself, as 
discussed in Section 2.2,
and the counterterms needed to cancel these 
violations in the infrared are totally independent of matter couplings and 
masses.

\section{Generalization to realistic non-perturbative matter}

The beauty of the second quantized version of 
Eq. (\ref{target-Cgrav}) is that it has a simple generalization 
to more general and realistic matter, 
 replacing the $\psi$ scalar functional integration by one over 
the Standard Model fields with Standard Model action, 
\begin{eqnarray}
\label{target-Cgravnmat}
{\cal Z} 
&=& \int [{\cal D} h_{\mu \nu} ~{\cal D}{\rm ghosts}]_{_{ \ell}} ~ e^{- S_{grav} - S_{c.t.} - 
\sum_{N} \sum_{b \in {\cal S}} \epsilon(b) 
\ln \{ \int_{\psi'(\partial b_N) =0} 
 {\cal D} \psi'_{SM} e^{-S_{SM}} \} }  \nonumber \\
&~& ~ ~ \times
\int 
 {\cal D} \psi_{SM} ~e^{-S_{SM}}.
\end{eqnarray}

 This effective 
description makes sense even when non-perturbative matter effects are 
important, such 
as in QCD. As long as there are no light  scalars 
$\ll 1/\ell$ with 
large non-derivative couplings in the matter sector we do not need 
the $\psi \rightarrow - \psi$ symmetry to repeat our analysis that 
the stable size of the cosmological constant is order $1/\ell^4$. 
That symmetry only helped to eliminate local terms sensitive to heavy masses 
from arising on 
the boundaries of blocks, $b_N$, upon integrating out heavy matter. 
(Of course here,
 heavy means in the sense of the physical spectrum, so for example 
the proton or pion 
is heavy even though one of their constituents is the ``massless''
 gluon.) However, by power-counting there are no such operators with 
spinor or vector light matter that preserve enough spacetime and gauge 
symmetry to be induced on 
the boundaries and have couplings of positive dimension (which could then 
be set by heavy masses). 

Once $S_{c.t.}$ is tuned to recover
 relativistic and BRST 
invariance in the gravitational couplings in the infrared,
Eq. (\ref{target-Cgravnmat}), interpreted either as a minimal effective 
description (I) or as a fat gravity toy model (II), 
satisfies the checklist from 
the introduction (i -- vii).  In the non-gravitational limit the matter sector 
has exact relativistic invariance. 

Eq. (\ref{target-Cgravnmat}) is essentially our final formulation, 
to be interpreted as an effective coarse-grained description of Nature, 
or as a fat gravity toy model. It must be confessed that we have not 
derived Eq. (\ref{target-Cgravnmat}) in its full generality, but rather we 
have explicitly done the special case of perturbative scalar matter and 
then simply jumped to the natural generalization. Hopefully we are still on 
the right track.

\section{Multiple Matter Vacua}

In this section, we will be putting our faith in Eq. (\ref{target-Cgravnmat})
interpreted as a fat gravity toy model. In particular, note that there is an 
important feature, namely that there appear to be
 two inequivalent ways we 
could try to add a purely classical cosmological constant term to 
Eq. (\ref{target-Cgravnmat}). One is to add an arbitrary constant 
vacuum energy density to the Standard Model action. It is straightforward to 
see that this way fails to modify the physical cosmological constant, 
since $S_{SM}$ appears in two places in Eq. (\ref{target-Cgravnmat}) in just 
such a way that a constant vacuum energy cancels out. The second 
way is to simply add a cosmological constant to $S_{grav}$. This of course, 
works. However, the decision we made in setting up the fat gravity toy model 
is that we imagine pure matterless fat gravity as unable to produce 
cosmological constant larger than $1/\ell^4$, that is, $S_{grav}$ contains 
a cosmological constant of at most $1/\ell^4$. These observations will be 
important in what follows. Of course, it is possible that fat gravity 
operates in a different way in Nature, but we will at least be able to examine 
one consistent possibility below.

A useful probe of any resolution of the cosmological constant problem 
is to imagine how things work when the matter sector contains long-lived 
metastable vacua in addition to the true vacuum. 
One way of phrasing the cosmological constant problem is that in the absence 
of gravity one can identify a Poincare-invariant matter vacuum and the 
problem  is how to maintain this invariance once gravity is ``added'', 
at least approximately over large spacetime regions. Metastable 
matter vacua in the absence of gravity can be approximately Poincare-invariant
over large regions of spacetime, so the question is whether whatever 
mechanism  one is considering for dealing with the cosmological constant 
suppresses it in the metastable phase as well as in the true vacuum.
If this is the case it would appear to make inflation impossible. This 
question is useful because it appears in standard effective field theory that 
if a metastable phase has very small cosmological constant, then the 
true vacuum must have lower energy density, which translates into a 
negative cosmological constant. If it is the true vacuum that has 
zero cosmological constant then the metastable vacuum must have higher 
energy density, translating into positive cosmological constant. 
It appears inconsistent for any mechanism to
eliminate the cosmological constant in all matter vacua, so it is 
interesting to  check in any proposal how the decision is made 
 as to which 
vacuum's cosmological constant to suppress, or whether in some 
mysterious way our standard expectations are violated. 

Here we will argue that in 
the toy model of fat Euclidean gravity given above,
the answer is that it is the true vacuum's cosmological constant that is 
suppressed, while metastable vacua have positive cosmological constant. That 
is metastable vacua are in an inflationary phase.

For simplicity, let us consider the case of (gravity coupled to) 
a single scalar 
matter field, $\psi$, with a double-well potential $V(\psi)$ with two 
inequivalent local minima, $V(\psi_t) < V(\psi_f)$. 
For this simple situation we can use  Eq. 
(\ref{target-Cgrav}) as a stripped down
 version of Eq. (\ref{target-Cgravnmat}).
We 
can expand about either the true vacuum at $\psi_t$ or the false one at 
$\psi_f$. We assume that about either vacuum the physical mass is large, 
$m_{\psi} \gg 1/\ell$. 
Since the decay of the false vacuum is a non-perturbative 
process, it will not appear at any order in the perturbative expansion 
about $\psi_f$. This will simplify things since we are not interested here 
in the decay but what happens in the long period (in real Lorentzian time) 
before the decay. Perturbation theory applied to the matter interactions 
 allows us to focus on this. 

In the absence of gravity, or with fixed background metric, the partition 
functional has the form, 
\begin{equation}
{\cal Z} = \int {\cal D} \psi e^{-S}.
\end{equation}
We approximate this integral by expanding about the minima of $S$, 
\begin{equation}
{\cal Z} \sim \int_{\psi \sim \psi_t} {\cal D} \psi e^{-S} + 
\int_{\psi \sim \psi_f} {\cal D} \psi e^{-S}, 
\end{equation}
where the action in the first term is taken to be expanded perturbatively 
about the Gaussian approximation centered on $\psi_t$, and the 
action in the second term is taken to be expanded perturbatively 
about the Gaussian approximation centered on $\psi_f$. That is, 
the partition function is the sum of the perturbative expansions about the 
two possible vacua. 
If we are interested in the theory around the false vacuum, $\psi_f$, then 
we only keep sources in the second term and the first integral is just some 
constant. If we think of $e^{-S}$ as a relative 
probability in the statistical 
interpretation of Euclidean field theory, then the first integral 
drops out of {\it conditional} probabilities, conditional on being in the 
vicinity of $\psi_f$. Similarly if we are interested in the true vacuum, 
$\psi_t$, we only keep sources in the first term and the second integral
drops out of conditional probabilities, conditional on being in the 
vicinity of $\psi_t$. 

Now let us turn gravity back on. Eq. (\ref{graveffS}) no longer automatically 
applies, we must re-think it starting from the fat gravity partition 
functional, Eq. (\ref{target-Cgrav}). Suppressing all details of gauge 
fixing, BRST violation and related counterterms, 
\begin{eqnarray}
\label{fatgrav}
{\cal Z}_{fat ~gravity} 
&=& \int [{\cal D} g_{\mu \nu}]_{_{ \ell}} ~e^{- S_{grav} - 
\sum_{N} \sum_{b \in {\cal S}} \epsilon(b) 
\ln \{ \int_{\psi'(\partial b_N) =0} 
 {\cal D} \psi' e^{-S_{matter}} \}}  
\int 
 {\cal D} \psi e^{-S_{matter}} \nonumber \\
&=& \int [{\cal D} g_{\mu \nu}]_{_{ \ell}} ~e^{- S_{grav} - 
\sum_{N} \sum_{b \in {\cal S}} \epsilon(b) 
\ln \{ \int_{\psi'(\partial b_N) =0} 
 {\cal D} \psi' e^{-S_{matter}} \}}  \nonumber \\
&~& ~ ~ ~ ~ \times \{\int_{\psi \sim \psi_t} 
 {\cal D} \psi e^{-S_{matter}} + \int_{\psi \sim \psi_t} 
 {\cal D} \psi e^{-S_{matter}} \},
\end{eqnarray}
where in the second line we have expanded the 
matter integral containing sources about the two possible vacua. 
The sources are in the $\psi_t$ ($\psi_f$)
 portion if 
we are interested in the true (false) vacuum, the non-source term dropping 
out of conditional probabilities as before. However, in either case all of the
sourceless block-wise functional integral appears in the gravitational 
Boltzman weight,  
exp($-\sum_{N} \sum_{b \in S} \epsilon_b 
\ln \{ \int_{\psi(\partial b_N) =0} 
 {\cal D} \psi e^{-S_{matter}}  \}$).    In this term 
the expansion about the true vacuum (minimal action) 
 dominates exponentially over 
 any false contribution.  
Integrating out the heavy matter about the true vacuum 
again gives local terms 
in the form of local bulk and boundary lagrangians. By Eqs. (\ref{bulk}) 
and (\ref{bdry}) the boundary terms cancel and the bulk terms add up to 
a local gravitational counter-term subtracting the vacuum energy at the 
{\it true} minimum. Thus whether one is expanding the correlators of the 
fat gravity theory about the true vacuum or the false vacuum, the 
 fat gravity effect is to subtract the true vacuum energy up to 
$\sim 1/\ell^4$ gravitational corrections. Thus the true vacuum has small 
cosmological constant while the false vacuum has large positive cosmological 
constant. 

\section{``Quenched Gravity''}

The results derived in this paper can be presented in a simple, 
manifestly general coordinate invariant prescription, valid either in 
Euclidean or Lorentzian spacetime. This prescription, which I will call 
``quenched gravity'' satisfies (i -- vii), and was discussed in Ref. 
\cite{old}. However, quenched gravity is stated directly in terms of the 
amplitudes and does not, at first sight, follow from a path integral in a simple 
way. 
To that extent, it may have seemed rather ad hoc, were it not for the 
preceding sections, which show how it arises naturally from a 
suppression of small vacuum loops in the path integral.

Quenched gravity applies when expanding about some matter (metastable) 
vacuum, with gravitational momentum transfers restricted to be below 
$1/\ell$, for some $\ell$. The general amplitudes are constructed in the 
usual way from tree diagrams, whose vertices are taken from the quantum 
1PI effective action. This can be decomposed as 
\begin{equation}
\Gamma_{eff} = \Gamma_{grav}[g_{\mu \nu}] + \Gamma_{matter}[g_{\mu \nu}, \psi],
\end{equation}
where all the vertices in the second term contains some matter lines. The 
rule is to compute $\Gamma_{matter}$ in exactly the standard way, 
but $\Gamma_{grav}$ as follows. First work in the absence of gravity 
about the matter vacuum of choice, and match to an effective Lagrangian 
valid below $1/\ell$, by integrating out heavier physics. For example, 
in the realistic case, if $1/\ell \sim 10^{-3}$ eV, all hadronic physics 
should be integrated out, because ``heavy'' refers to the physical spectrum
 not to ``massless'' gluons. This effective Lagrangian is then coupled to 
gravity below $1/\ell$ and used to compute $\Gamma_{grav}[g_{\mu \nu}]$, 
using dimensional regularization and minimal subtraction. 

The size of the cosmological constant in $\Gamma_{grav}[g_{\mu \nu}]$ is 
then obviously of order $1/\ell^4$ or less. We now add to 
$\Gamma_{grav}[g_{\mu \nu}]$ two local cosmological constant contributions. 
One is at the least of order $1/\ell^4$ 
with unknown  coefficient, representing 
the vacuum energy contributions of new gravitational degrees of freedom 
which are at least of mass $1/\ell$, or if one prefers, it represents 
the quartic sensitivity to the $1/\ell$ cutoff of the low-energy 
effective field theory which is missed by dimensional regularization.
The second contribution is from the vacuum energy density difference between 
the vacuum about which one is expanding and the true vacuum. 

Of course if one is expanding about the true vacuum, quenched gravity 
yields a natural size of the effective cosmological constant of order 
$1/\ell^4$, while describing gravitational momentum transfers below $1/\ell$ 
(and general matter momenta). If new gravity experiments probe shorter 
distances $\ell' < \ell$  and reveal only Newtonian behavior, then the 
 relativistic completion of General Relativity must continue to hold as a 
quantum effective field theory up to $1/\ell'$. That is, we must replace 
$\ell \rightarrow \ell'$ in our prescription. Now the natural size of the
cosmological constant is $1/\ell'^4$. If this is much larger than the 
dark energy density of the universe then the cosmological constant 
problem has returned, {\it but not until then}. In the realistic case we 
are not there yet.

\section{Conclusions}

This paper has taken a first quantized Euclidean  path to showing that a 
minimal description of the present experimental domain does not imply a 
cosmological constant fine-tuning problem. But if gravity described by 
effective quantum general relativity (that is, a point-like nature for the 
graviton) is extrapolated to distances much shorter than $100$ microns, the 
cosmological constant problem does emerge. Naturalness then suggests that 
new gravitational physics should be revealed in future short-distance 
gravitational tests. The fat gravity toy models in this paper give a concrete 
illustration of the qualitative features of such new physics: suppression 
of the cosmological constant (of the true vacuum, if there are metastable 
vacua) in conjunction with suppression of the short-distance gravitational 
force.

Given that new gravitational physics is close at hand, it would be very 
useful to identify, estimate and parametrize accessible precision effects 
at distances above the transition scale $\ell$, within effective 
descriptions such as discussed here and in Ref. \cite{old}. It also 
remains to search for a consistent stringy realization of fat gravity.

I suspect that there is a more elegant ``block-spin'' coarse-graining 
than presented here,
based on the Regge calculus \cite{regge} which manifestly retains a discrete 
subgroup of general coordinate invariance. 
However, several technical issues remain to be worked 
through before this is clear.

\section*{Acknowledgements}

I am grateful to Nima Arkani-Hamed, Savas Dimopoulos, Markus Luty
and Joe Polchinski for emphasizing to me some of the issues that I have 
tried to worry about in this work.
This research was supported  by 
NSF Grant P420D3620434350.

\appendix 

\section{Appendix}

In this appendix, we will work for convenience in units in which $\ell = 1$. 
Most of the results on webs and blocks used in the 
paper will be proven by induction on the number of Euclidean spacetime 
dimensions, $d$. Of course in the end we will be interested in $d = 4$.
The results for which we use induction are restated for general $d$ as 
follows:

In $d$ dimensions, there exists a subset, $C$,  of connected vacuum webs, 
a set, ${\cal S}$,
 of composite blocks (composed of unit lattice cells), $b$,  
which are contained in the region $[0, 2]^d$, 
and a binary-valued function on ${\cal S}$, 
$\epsilon(b) = \pm 1, b \in {\cal S}$ such that

(a) \begin{equation} 
\int [{\cal D} w]_C ... = \sum_N \sum_{b \in {\cal S}} \epsilon(b) 
\int [{\cal D} w]_{b_N} ... ~.
\end{equation}

(b) If the diameter of a vacuum web is $< 1$ then it must be an element of 
$C$.

(c) \begin{equation}
\sum_N \sum_{b \in {\cal S}} \epsilon(b) \int_{b_N} d^dx ... = \int d^dx... ~.
\end{equation}

(d) \begin{equation}
\sum_N \sum_{b \in {\cal S}} \epsilon(b) \int_{\partial 
b_N} d^{d-1}x ... = 0.
\end{equation}

We will first prove this for the case where the metric is exactly flat, as 
is the case in the $\phi$ model without real gravity, and then 
briefly discuss real 
gravity afterwards.

We begin by giving a proof for a single spacetime dimension, $d = 1$. 
Here we define $C$ to be the connected vacuum webs which are entirely 
contained within some composite interval (one-dimensional block) of length 
$2$, that is $[N_1, N_1+2]$ for some integer $N_1$. 
${\cal S}$ consists of two blocks, 
$[0, 2]$ with $\epsilon = +1$,  and $[0, 1]$ with $\epsilon = -1$. Now let 
us prove (a -- d). 

(a) Naively, one might think that
\begin{equation}
\int [{\cal D} w]_C ... = \sum_{N_1}  \int [{\cal D} w]_{[N_1, N_1 +2]}... ~.
\end{equation}
But this is wrong because some webs in C might sit inside two 
overlapping length-$2$ intervals, $[N_1-1, N_1+1]$ and $[N_1, N_1+2]$, 
so that they get counted twice on the right-hand side but just once on the 
left. But in precisely these cases, the web must be within the intersection, 
$[N_1, N_1 + 1]$. Thus the double-counting is removed by subtracting the 
integral over webs which fit into such a unit interval. That is, 
 \begin{equation}
\int [{\cal D} w]_C ... = \sum_{N_1}  
(\int [{\cal D} w]_{[N_1, N_1 +2]} - \int [{\cal D} w]_{[N_1, N_1 +1]})... ~,
\end{equation}
which is just what we had to prove.

(b) If the diameter of a web is less than $1$ then clearly the web must 
be contained in some length-$2$ interval of the form $[N_1, N_1+2]$, 
and hence is in $C$.

(c) We have, 
\begin{eqnarray}
\sum_{N_1} (\int^{N_1 + 2}_{N_1} dx_1 - \int^{N_1 + 1}_{N_1} dx_1) ... 
&=& \sum_{N_1} \int^{N_1 + 2}_{N_1 +1} dx_1 ... \nonumber \\
&=& \int_{-\infty}^{\infty} dx_1 ... ~.
\end{eqnarray}

(d) If we have some function of $1$-dimensional spacetime, $f(x_1)$, 
then clearly 
\begin{equation}
\sum_{N_1} (f(N_1 + 2) + f(N_1) - f(N_1 +1) - f(N_1)) = 0.
\end{equation}

Now let us make the inductive assumption that our results are true for 
$d$ Euclidean spacetime dimensions. We will prove that they must then 
also be true in $(d + 1)$ dimensions. In order to not get confused between 
$d$-dimensional and $(d + 1)$-dimensional objects we will denote 
the latter using barred symbols, 
$\overline{C}, \overline{{\cal S}}, \overline{b}, \overline{\epsilon}, ...~$. 

We define $\overline{C}$ to contain any vacuum web $\overline{w}$ which is 
contained in some thickness-$2$ {\it slab} of the form, $N_{d+1} 
\leq x_{d+1} \leq N_{d+1} +2$, {\it and} where the projection of 
$\overline{w}$ to $d$ dimensions, $\overline{w}_P \in C$. 
Here ``projection'' has the straightforward interpretation that if one 
throws away the $d+1$-th coordinate information for a $(d+1)$-dimensional 
web, one is left with a $d$-dimensional web. We define $\overline{{\cal S}}$
 as 
consisting of composite blocks $\overline{b} \equiv b \times [0, 2], b \in 
{\cal S}$ 
with $\overline{\epsilon}(\overline{b}) \equiv  \epsilon(b)$, 
and $\overline{b} \equiv b \times [0, 1], b \in S$ 
with $\overline{\epsilon}(\overline{b}) \equiv  - \epsilon(b)$. Obviously 
the $\overline{b}$ are contained within $[0,2]^{d+1}$ since the 
$b$ are contained within $[0,2]^{d}$. 

($\overline{a}$) We have, 
\begin{eqnarray} 
&\sum_{\overline{N}}& \sum_{\overline{b} \in \overline{{\cal S}}} 
\overline{\epsilon}(\overline{b}) 
\int [{\cal D} \overline{w}]_{\overline{b}_{\overline{N}}} ...
= \sum_{N_{d+1}} \sum_N \sum_{b \in {\cal S}} \epsilon(b) 
(\int [{\cal D} \overline{w}]_{b_N \times [N_{d+1}, N_{d+1} +2]} 
-\int [{\cal D} \overline{w}]_{b_N \times [N_{d+1}, N_{d+1} +1]})... 
\nonumber \\
&=& \sum_{N_{d+1}} \sum_N \sum_{b \in S} \epsilon(b) 
\int [{\cal D} w]_{b_N} (\int [{\cal D} x_{d+1}(w)]_{[N_{d+1}, N_{d+1} +2]} 
- \int [{\cal D} x_{d+1}(w)]_{[N_{d+1}, N_{d+1} +1]}) ... \nonumber \\
&=_{{\rm by} (a)}& \sum_{N_{d+1}} \int [{\cal D} w]_C 
(\int [{\cal D} x_{d+1}(w)]_{[N_{d+1}, N_{d+1} +2]} 
- \int [{\cal D} x_{d+1}(w)]_{[N_{d+1}, N_{d+1} +1]}) ... \nonumber \\
&=& \int [{\cal D} \overline{w}]_{\overline{C}} ... ~.
\end{eqnarray}
In the second line we have used the fact that a $(d+1)$-dimensional 
web, $\overline{w}$, is specified by lifting a $d$-dimensional web, $w$, 
by supplementing it with a function, $x_{d+1}(w)$, giving the ``elevation''
 of the
web in the $(d+1)$-th dimension. In the third line we have used the 
inductive assumption (a) in $d$ dimensions. In the fourth line we are 
using the obvious fact that the projection of the lift takes you back to the 
starting web, $\overline{w}_P = w$, and the definition of $\overline{C}$ 
requiring that webs fit into thickness-$2$ slabs. 
 The possible 
non-uniqueness of such a slab for a given web in $\overline{C}$
and the related issue of double-counting of webs is in exact analogy to the 
$d=1$ case, that is the second term on the fourth line precisely removes the 
doubly-counted webs from the first term. 

($\overline{b}$) If the diameter of $\overline{w}$ is less than $\ell$, 
then it must certainly lie in some thickness-$2$ slab, 
$N_{d+1} 
\leq x_{d+1} \leq N_{d+1} +2$. Furthermore, the diameter of 
$\overline{w}_P$ must be at most that of $\overline{w}$ because of the 
Euclidean metric. Therefore by the inductive assumption (b) and the 
definition of $\overline{C}$, 
$\overline{w} \in \overline{C}$.

($\overline{c}$) We have,
\begin{eqnarray}
\sum_{\overline{N}} \sum_{\overline{b} \in \overline{{\cal S}}} 
\overline{\epsilon}(\overline{b}) \int_{\overline{b}_{\overline{N}}} d^{d+1}
\overline{x} 
... &=& \sum_{N_{d+1}} \sum_N \sum_{b \in S} \epsilon(b) 
\int_{b_N} d^dx (\int_{N_{d+1}}^{N_{d+1} +2} d x_{d+1} - 
\int_{N_{d+1}}^{N_{d+1} +1} d x_{d+1}) ... \nonumber \\
&=_{{\rm by} (c)}& \sum_{N_{d+1}} \int d^dx \int_{N_{d+1}+ 1}^{N_{d+1} +2} d x_{d+1} ...
\nonumber \\
&=& \int d^{d+1}
\overline{x} ... ~.
\end{eqnarray}

($\overline{d}$) We have,
\begin{eqnarray}
\sum_{\overline{N}} \sum_{\overline{b} \in \overline{{\cal S}}} 
\overline{\epsilon}(\overline{b}) \int_{\partial 
\overline{b}_{\overline{N}}} d^{d}
\overline{x} 
... &=& \sum_{N_{d+1}} \sum_N \sum_{b \in S} \epsilon(b) 
(\int_{b_N, x_{d+1} = N_{d+1}+2} d^dx + \int_{b_N, x_{d+1} = N_{d+1}} d^dx
\nonumber \\
&~& ~ ~ + \int_{\partial b_N} d^{d-1} x \int_{N_{d+1}}^{N_{d+1}+2} d x_{d+1} 
- 
\int_{b_N, x_{d+1} = N_{d+1}+1} d^dx \nonumber \\
&~& ~ ~ 
- \int_{b_N, x_{d+1} = N_{d+1}} d^dx
- \int_{\partial b_N} d^{d-1} x \int_{N_{d+1}}^{N_{d+1}+1} d x_{d+1}) ... 
\nonumber \\
&=_{{\rm by} (d)}& \sum_{N_{d+1}} \sum_N \sum_{b \in S} \epsilon(b) 
(\int_{b_N, x_{d+1} = N_{d+1}+2} d^dx - \int_{b_N, x_{d+1} = N_{d+1}+1} d^dx) 
... \nonumber \\
&=& 0.
\end{eqnarray}

This completes the proof by induction. 

There is one more result which we need and can now prove directly, without induction on the number of dimensions, 
namely that 
any web with diameter greater than $4$ is not contained in $C$. 
We have (directly in $4$ dimensions), using the result (a), that any web in 
$C$ is certainly contained in the region $[0,2]^4$ or one of its translations.
But the greatest distance between any two points within 
$[0,2]^4$ (or its translations) is obviously 
$4 = \sqrt{2^2 + 2^2 +2^2 + 2^2}$. Therefore any web in C cannot
 have diameter greater than $4$.

Now, all the above proofs have assumed a flat Euclidean metric. In the case 
of real gravity (as opposed to $\phi$) this does not hold. The issue of
the metric is only relevant to those results giving lower and upper bounds on 
the diameter of class C webs. These in turn are used in the paper to show 
that all vacuum webs in class D have diameter $> \ell$ and therefore 
heavy vacuum loops in class D are exponentially suppressed, and that 
class C webs have small enough diameter to be matched to local 
operators. These results continue to hold in the case of real gravity  
since we are only 
considering perturbative gravity in this paper, the metric is always being 
expanded about, and is close to, flat space.

\end{document}